\documentclass[12]{article}
\usepackage{graphicx}
\usepackage[top=2cm, bottom=3cm, left=3cm, right=3cm]{geometry}

\title{Secondary cosmic ray nuclei in the light of the Single Source Model and comparison with the recent AMS-02 data}
\author{$^{*}$A.D. Erlykin $^{1,2}$, A.W. Wolfendale $^{2}$\\
$(1)$ P N Lebedev Physical Institute, Moscow 119991, Russia.\\
$(2)$ Physics Department, Durham University, Durham, DH1 3LE, UK\\}

\begin{document}

\maketitle

\footnote{$^{*}$Corresponding author: tel +74991358737 \\ 
 E-mail address: erlykin@sci.lebedev.ru}

\begin{abstract}
Evidence for a local 'Single Source'  of cosmic rays is amassing by way of the recent precise  measurements of various
 cosmic ray energy spectra from the AMS-02 instrument. To observations of individual cosmic ray nuclei, electrons,
positrons and antiprotons must now be added the determination of the boron-to-carbon ratio and the energy spectrum of 
lithium to 2000 GV with high precision. Our analysis leads us to claim that, with certain assumptions about 
propagation in the Galaxy, the results confirm our arguments regarding the presence of a local single source, 
perhaps, a supernova remnant (SNR). An attempt is made to determine some of the properties of this SNR and its 
progenitor star.  
\end{abstract}

\section{Introduction}
Nuclei of Li, Be and B in cosmic rays (CR) are certainly of secondary origin since their abundance in the 
solar system is negligibly small. The term 'secondary' (S) is introduced to distinguish them from their parents, 
which are called 'primary' (P) and are heavier than the secondaries. Collisions of primaries with atoms of the 
interstellar medium  (ISM) result in their fragmentation and give rise to the production of Li, Be, B and other 
fragments. The S/P ratio had been widely used to study CR propagation. Besides the boron-to-carbon (B/C) ratio 
other ratios such as F/Ne, (Sc+Ti+V+Cr)/Fe have been used. In particular, the mean free path (grammage) for the 
escape from the Galaxy $\lambda_{esc}$ and its energy dependence were estimated. The radioactive isotopes such as 
$Be^{10}, Al^{26}, Cl^{36}, Mn^{57}$ were also used to estimate the lifetime of CR nuclei in the Galaxy $\tau_{esc}$. In 
this paper 
we concentrate on the analysis of the recent high precision measurements of the B/C ratio and the Li rigidity 
spectrum presented at CERN by the AMS-02 collaboration (Oliva for AMS-02 coll., 2015; Derome for the AMS-02 coll., 
2015). The data were analysed in several earlier papers viz. by Cowsik R. et al., 2014, Tomassetti, 2015a etc. In a 
set of our own recent papers we analysed the AMS-02 data on the proton and helium energy spectra, positron and 
antiproton fractions (Erlykin and Wolfendale, 2013, 2015a, 2015b) in terms of possible contributions of the local 
source. In the present paper we examine the possibility of reconciling the B/C ratio and the Li spectrum measured by 
the AMS-02 experiment with our Single Source Model (SSM).

Arguments favouring a local single recent CR source for particles below some tens of TV are well known (eg. Erlykin
 and Wolfendale, 1997, 2005), but the case is not yet fully proven. Recent precise AMS-02 measurements of 
protons (Aguilar et al., 2015) and helium (Haino for the AMS-02 coll., 2015) spectra, the positron and electron 
spectra and fractions (Accardo et al., 2014; Aguilar et al., 2014) and antiprotons (Kounine for the AMS coll., 
2015) have been explained by us and others in terms of a local, recent single source (Erlykin and Wolfendale, 
2013, 2015a, 2015b; Tomassetti, 2015c). The latest AMS-02 results concerning the
 B/C ratio (Oliva for the AMS-02 coll., 2015) and Li rigidity spectrum (Derome for the AMS coll., 2015) give the 
opportunity of a further check on the hypothesis.          

B/C ratio determination has a long history and  has led to a determination of the mean CR lifetime $\tau(R)$ as a 
function of rigidity $R$ (see the bibliography in Gaisser, 1990). We have consistently adopted the form 
$\tau(R)=4\cdot 10^7y$ at $R\approx 1 GV$ with an energy dependence of $\tau$ of the form $R^{-0.5}$ (Erlykin and 
Wolfendale, 2001). At $R = 100GV$, say, the mean lifetime against escape from the Galaxy is thus $4\cdot 10^6y$, a 
time interval greatly in excess of the ages of local known CR sources, supposed to be supernova remnants (SNR).

The B- nuclei, and other secondaries, are produced by interactions in the ISM both within the remnants (for the usual
 SNR sources) and in the ISM in general. The secondaries travel through the ISM in the same manner as the primaries 
and, for the usual 'leaky box' model of the Galaxy, the lifetime factor appears once in the expression for the 
intensity of the primaries, but twice for the secondaries. That is why the B/C ratio falls with rising rigidity.
\section{The results and the analysis} 
\subsection{Estimates of the escape length}
\subsubsection{Rigidity dependence of the grammage propagated by cosmic ray particles before they escape from the 
Galaxy}
 For the stable isotopes the diffusion model of the CR propagation in the Galaxy is equivalent to the 
'leaky box' model. In the framework of this model the stationary CR secondary to primary S/P ratio is desribed as 
(Gaisser, 1990) 
\begin{equation}
S/P = \frac{\lambda_{esc}}{\lambda_{ps}(1+\frac{\lambda_{esc}}{\lambda_s})}
\end{equation} 
Here $\lambda_{esc}$ is the escape length, or the grammage, propagated by cosmic rays before they escape from the 
Galaxy, $\lambda_{ps}$ is the cross section for the fragmentation of primary P nuclei into secondary S nuclei and 
$\lambda_s$ is the cross section for the inelastic collisions of S nuclei with ISM nuclei. At rigidities above 
1 GV $\lambda_{ps}$ and $\lambda_s$ are approximately constant and only $\lambda_{esc}$ is energy dependent. Since 
at fixed rigidity the escape is the same for all particles the energy dependence of $\lambda_{esc}$ can be 
derived from a comparison of the emergent spectrum of protons with the observed proton spectrum. 
The emergent spectrum of protons accelerated by the shocks in supernova explosions has been calculated by Erlykin 
and Wolfendale (2001). In the energy range of 1 GeV - 1 PeV it has a power law energy behaviour with the differential
 energy index of $\gamma = 2.15\pm 0.02$. Such a value is common for all models involving Fermi-acceleration. 
Experimental measurements of the gamma-ray energy spectra for several identified SNR show a rather broad range of the
indices $\gamma$ from 1.95 to 2.5 (Rieger et al., 2013). Our adopted value of $\gamma = 2.15$ is right in the middle 
of this range. The proton spectrum observed in the AMS-02 experiment is fitted by two power laws smoothly connected 
with each other. In the 30-200 GeV energy range coincident with the
corresponding rigidity the observed spectrum is steeper than the emergent one and has the power index of 
$\gamma = 2.849\pm 0.002$ (Aguilar et al.,2015). The difference in the indices is due to escape so that the 
rigidity dependence of $\lambda_{esc}$ can be fitted as $\lambda_{esc} = \lambda_0(R/R_{min})^{-\delta}$ with
  $\delta = \Delta\gamma = 0.70\pm 0.02$. Here {\em R} is the rigidity and $\lambda_0$ is an escape length at low 
energy, which we will discuss later.

Another approach to the rigidity dependence of $\lambda_{esc}$ can be connected with energetics arguments. The 
energy density of cosmic rays produced by SNR can be estimated as $Q = \frac{E_{CR}\tau_{esc}}{VT}$ where $E_{CR}$ 
is the total energy injected by SNR into CR, $V = 2\pi R_G^2H_G$ is the volume of the Galaxy, $\tau_{esc}$ is the 
mean lifetime of CR in the Galaxy and {\em T} is the mean period between SN explosions. If $E_{CR} = 10^{50}$
erg,  $R_G, H_G$ are the radius and half-height of the galactic disc equal to 15 kpc and 1 kpc respectively, 
$\tau_{esc} = 4\cdot 10^7$y and $T = 50$y, then the energy input rate and the energy density of CR in the Galaxy 
are $1.0\cdot 10^{-15} eVcm{^{-3}}s^{-1}$ and $1.3eVcm^{-3}$ respectively, which are reasonable values estimated 
still in the sixties by Ginzburg and Syrovatsky, (1964).  

If the emergent spectrum has the power law behaviour $I_{CR}=A(E/E_{min})^{-\gamma}$ then the total CR energy 
density above $E_{min}$ is $\rho_E = \frac{4\pi AE_{min}^2}{c(\gamma-2)}$ with $c$ as the speed of light. 
 We compare estimated and observed 
energy density at energies above 45 GeV chosen by the AMS-02 collaboration as the fitting parameter. If in our 
emergent power law spectrum $E_{min} = 1 GeV$, then above 45 GeV CR carry $45^{-(\gamma-2)}=0.56$ fraction  
of the total energy. The power law part of the proton spectrum above 45 GeV ($\gamma=2.849$) observed by AMS-02 
collaboration contains 0.045 eVcm$^{-3}$. In order to get this energy content the lifetime of CR above 45 GeV 
should be $\frac{0.045}{1.0\cdot 10^{-15}\cdot 0.56} = 8\cdot 10^{13}$s = $2.54\cdot 10^6$y. Since the CR lifetime
being $4\cdot10^7y$ at 1 GeV diminishes to $2.54\cdot 10^6y$ at 45 GeV then the lifetime has to fall as 
$(\frac{E}{E_{min}})^{-\delta}$ with $\delta = 0.72$ which agrees well with the estimate of 
$\Delta\gamma = 0.70$ made from the difference of the power indices of the observed and emergent proton spectra.

We understand all the uncerainties incorporated into these estimates and consider all numerical estimates
 described above only as an illustration. However, we are sure that the lifetime of primaries in the Galaxy has 
to decrease with the energy faster than that with the power index of 0.3-0.6 derived usually from the energy 
dependence of boron-to-carbon ratio and with the index 0.5, as used by us in the past. 
 
\subsection{The estimate of the absolute value of $\lambda_{esc}$}
The preliminary data on the B/C ratio presented by AMS-02 collaboration (Aguilar et al., 2015) demonstrate the 
power law behaviour only at rigidities above ~20 GV. With decreasing rigidity below ~20 GV the B/C ratio 
becomes flatter and approaches a constant at 2-3 GV. We think that solar modulation is not responsible for 
this flattening, because for the fixed rigidity it should reduce boron and carbon fluxes by the same factor.
The B/C ratio is described by the expression (1) in which symbols S and P should be replaced by B and C, but C 
should include also all parent nuclei heavier than C. As has been already mentioned the mean free paths 
$\lambda_{BC}$ and $\lambda_B$ are nearly constant at energies above 1 GeV so that only the term 
$(1+\frac{\lambda_{esc}}{\lambda_B})$ can distort the power law behaviour of B/C at low energies, where 
$\lambda_{esc}$ is comparable with $\lambda_B$. In order to make numerical estimate we used cross sections 
$\sigma_{BC}$ and $\sigma_B$ from Silberberg and Tsao, (1973) updated recently by Tomassetti, (2015b). In the 
calculations the production of both $^{10}B$ and $^{11}B$ isotopes was taken into account and the cross sections
of the particular reactions were weighted with the abundance of the parent nuclei taken from Sihver et al.(1993).
The calculated fragmentation length $\lambda_{BC}$ and the inelastic interaction length $\lambda_B$ were obtained 
as $\lambda_{BC} = 161 gcm^{-2}$ and $\lambda_B = 85.5 gcm^{-2}$. At a rigidity of 2 GV the B/C ratio is 0.31 
(Oliva for the AMS-02 call., 2015) and using this ratio with the parameters $\lambda_{BC}$ and $\lambda_B$ in the 
expression (1) we obtain $\lambda_{esc}(2GV) = 126 gcm^{-2}$. Taking this value as the $\lambda_0$ parameter we conclude
 that
\begin{equation}
\lambda_{esc}(R) = 126(\frac{R}{2GV})^{-0.70}
\end{equation}        
The grammage is the product of the mean gas density $\rho$ and the pathlength $\ell$ propagated during their lifetime 
by CR primaries in the region where they produced boron: $\lambda_{esc} = \rho v \tau_{esc}$ with $v$ as the velocity 
of the particle. Since measurements of radioactive isotopes give an estimate $\tau_{esc} \approx (18\pm 3)\cdot10^6y$ in the GV region (Connell et al.,
1997) then $\rho \approx 10^{-23}gcm^{-3}$ i.e. by an order of magnitude higher than the usually taken value of 
$10^{-24}gcm^{-3}$. To us it gives evidence that boron is produced mainly in the dense envelopes of SNR.     
 
\subsection{The Evidence for the Single Source}
\subsubsection{The B/C ratio}
The B/C ratio expected from the expression (1) is shown in Figure 1 by the full line denoted as 'BGRD' (background). 
\begin{figure}[htb]
\begin{center}
\includegraphics[width=8cm,height=15cm,angle=-90]{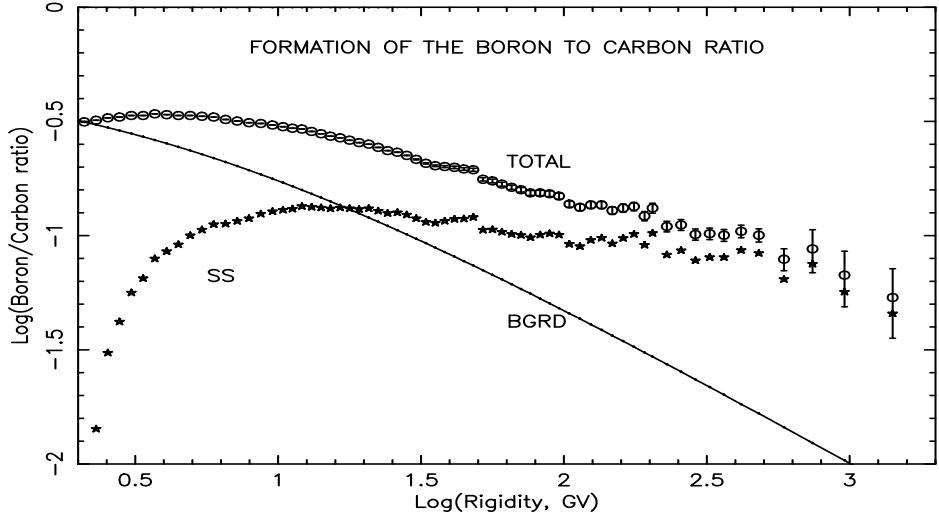}
\end{center}
\caption{\footnotesize The boron to carbon ratio as a function of the rigidity. Experimental poins are from Oliva,
 (2015), the full line - the 'background' ratio expected from the expression (1) and denoted as 'BGRD'. Stars 
indicate the difference between the AMS-02 experimental ratio and the BGRD expectation attributed to the contribution
 of the Single Source (SS).} 
\label{fig1}
\end{figure}
It is seen that in the region above several GV the experimental ratios are in excess of expectation. We 
attribute this excess to the contribution of the local SNR which we call 'the Single Source'. The expected B/C ratio
described by the expression (1) is supposed to be due to the background from the old and distant SNR. In fact, the
 difference between the two B/C ratios is not equivalent to the B/C ratio of one of these components. However, an
examination of the flattenings in the carbon and oxygen energy spectra (see Adriani et al., 2014) which presumably 
are due to contributions of the Single Source shows that this contribution is small. Therefore, the difference 
between the observed and expected B/C ratios can be referred to the $\frac{B_{ss}}{C_{bgrd}}$ ratio and its shape 
reflects the energy behaviour of the boron energy spectrum. Since it is expected that the Single Source is a local 
SNR then its energy spectrum is hard and its contribution to the ratio at GeV energies is negligibly small. The 
observed ratio in the GeV region is completely determined by the background and our determination of $\lambda_{esc}$ from the expression 
(1) is based on this assumption.
\subsubsection{The energy spectrum of Lithium}
At \'AMS-02 days at CERN\' the preliminary data on the rigidity spectrum of Lithium was presented by Derome and 
Ting, 2015). It is shown in Figure 2 in double-logatithmic scale.       
\begin{figure}[hptb]
\begin{center}
\includegraphics[width=8cm,height=15cm,angle=-90]{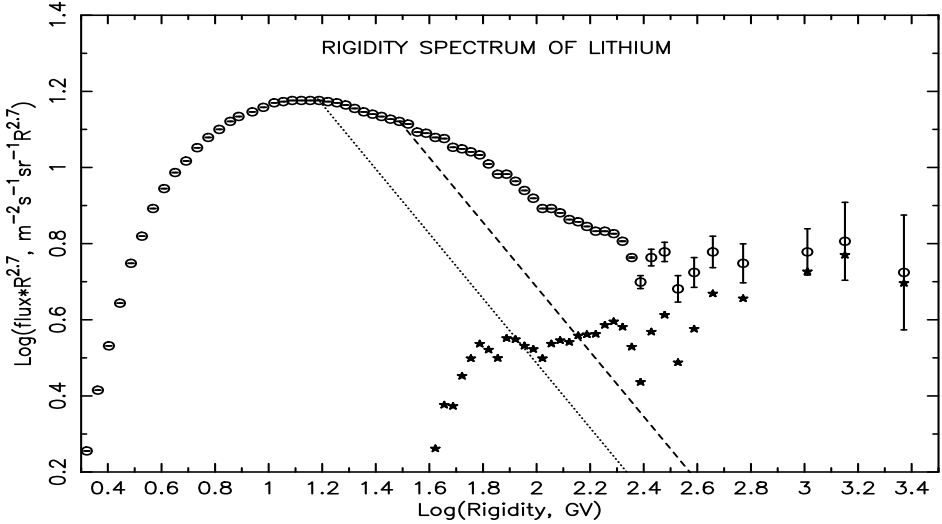}
\end{center}
\caption{\footnotesize The energy spectrum of lithium as a function of the rigidity. Experimental points are from 
Derome, (2015), dashed line - the 'background' with the exponent of the rigidity dependence $\delta = 0.7$ above 30 
GV (30 GV is taken as the 'joining point'), dotted line - the background of the same shape above 15 GV. The stars are
 the difference between the AMS-02 experimental data and the expected background (dashed line) attributed by us to 
the contribution of the Single Source (SS).} 
\label{fig2}
\end{figure}

Experimental points (open circles) are from Ting (2015). It is seen that the spectrum has an evident irregularity - 
it flattens above ~300GV like spectra of protons and other nuclei. The dashed line is the 'background expected from 
the assumption that Li parent nuclei (C,N,O etc.) being primaries have the same rigidity spectrum as protons above 
30GV, i.e. $\gamma = 2.85$ and the escape length line is determined by the expression (2). The stars are the 
difference between the experimental data and the background. The origin of the dotted line will be discussed later. 
   
\section{Discussion}
Boron and Lithium are secondary nuclei. It is interesting to compare them with the other secondary CR components - 
positrons and antiprotons Taking 200 GV as the datum rigidity we compare the enhancement factors $F$ = SS intensity/
BGRD intensity. We have derived $F$ for each component with the result:
 $F_{e^+} = 8, F_{\bar{p}/p} = 2.6, F_{B/C} = 2.7, F_{Li} = 1.3$. Errors are not quoted because unknown systematic errors 
are significant. For instance, if the starting rigidity for the steeper Li background spectrum is reduced from 30GV 
to 15 GV (dotted line in Figure 2) then the enhancement factor $F_{Li}$ increases from 1.3 to 2.7. Taking into 
account 
that the higher enhancement factor $F_{e^+}$ for positrons can be due to the contribution of a pulsar associated with 
SNR it can be said that all the other enhacements are of the same order of magnitude. This fact can be regarded as 
a hint that the excess of all secondaries is produced by the same Single Source, presumably a SNR. The large value of
$\lambda_{esc}$ indicates that the region where they are produced has relatively high density, most likely within the 
SNR envelope or in the nearby molecular cloud. The consequence of this indication is that the SNR should be the gamma 
 source since gamma quanta are also secondaries and should be copiously produced on the thick target. 

The question is whether the assumed SNR is an average one or it has special properties different from the average.
Here, attention can be drawn to the possible scenario in which it is pointed out that some young Type II SN 
produce high quantity of $^{11}B$. The value of $^{11}B/Fe$, for example, increases by between 10 and 100 for shocks of
velocity 6000kms$^{-1}$ compared with 4000kms$^{-1}$. Furthermore, for the same shock velocity the $^{11}B/Fe$ ratio 
increases by a factor of 300 in going from effective radius of $10^{13.66}cm$ to $10^{12.78}cm$ (Dearborn et al.,1989). 
Clearly, there is considerable sensitivity and our estimate of $F_{B/C}$ could be subject to change.

Turning to $F_{e^+}$ presumably most positrons are produced in $\pi^+ \rightarrow \mu^+ \rightarrow e^+$ decays and 
their numbers will depend on the location of interactions, the gas density there and the trapping by the ambient 
(compressed ?) magnetic field. Again, there is sensitivity to the conditions in the SNR.

Finally, with respect to $F_{\bar{p}/p}$, the situation is similar to that for $F_{e^+}$. The number of $\bar{p}$  will 
depend on the location of their parent's interactios as their subsequent acceleration. As it has been already 
mentioned the analysis of positron/electron and antiproton/proton ratios gives support to the significant 
contribution of the the local Single Source to these ratios (Erlykin and Wolfendale, 2013, 2015b). 

At present it seems more likely that the local Single Source is a roughly typical SNR. We cannot point to a 
paticular nearby SNR, but our guess is that Vela is too young and not seen with protons at GeV and TeV energies. Its
 contribution is seen in the sharp knee in the PeV region. The most likely candidates which produce an excess of 
secondary particles are Geminga and Monogem Ring SNR. Since the Monogem Ring is seen only in X rays then Geminga is  
preferrable as the Single Source since it is also an intensive gamma-ray source.   

A technique that is hoped to adopt in the future in order to elucidate the properties of the SN is to examine the SN 
models from the standpoint of determining the relative masses of nuclei in the stellar wind of the pre-cursor star. 
Comparison of the measured CR fluxes with those from the stellar wind should yield the mass of the progenitor star. A 
first order attempt has been made using the analysis of Arnett et al.(1989), which, related to SN 1987A, yields a mass
of $~15 M_{\odot}$ but further, more sophisticated analysis is needed.

A further on-going analysis relates to possible fine structure in the spectral data. We draw attention to possible 
small excesses over smooth spectra in the region of 200 GV for $e^+/e^-$ B/C and Li. Furthermore, there is a deficit in
the iron spectrum at the same rigidity in the iron intensity that shows up in the O/Fe and C/Fe (primaries , not 
secondaries, see Tomassetti, 2015c). This can be understood in terms of the natural variation in ISM variations from 
place to place, as well as the dependence of the iron yield in the pre-cursor stellar mass.    
\section{Conclusions}
The data on the B/C ratio and the Li rigidity spectrum presented by the AMS-02 collaboration at 'Days of AMS-02 at 
CERN' meeting (Oliva for the AMS-02 coll., 2015; Derome for the AMS-02 coll., 2015, Ting for the AMS-02 coll., 2015) 
are preliminary, therefore results 
of our analysis given in this paper are also preliminary. However, we think that most of these results will survive 
with time since the precision of the AMS-02 experimental data is extremely high even at the preliminary stage.

We think that the most convincing argument for the contribution of the Single Source to the B/C ratio and the Li 
rigidity spectrum is the difference between the relatively flat rigidity dependence of these secondaries 
($\delta \approx 0.33$) and the steep slope of the expected 'background' spectra ($\delta \approx 0.70$) expected 
from the comparison of the proton spectrum, measured by the AMS-02 collaboration ($\gamma_p = 2.85$) and our SNR 
emergent spectrum ($\gamma = 2.15$). All other arguments can be regarded as hints, although the approximate coincidence
of enhancement factors for antiprotons, B/C ratio and Lithium rigidity spectrum is encouraging. 

We appreciate that 
there are questions on the described scenario. Observations of gamma rays from some SNR indicate the emergent spectra 
of CR with the exponential index of $\gamma \approx 1.95-2.5$, which can be steeper than that adopted by us here 
as $\gamma = 2.15$ on the basis of the theory of Fermi-type acceleration in SNR. However, the number of such 
observations is still poor to disprove the SNR origin of most CR. Another interesting question relates to the 
expected rapid rise of the CR anisotropy with the rising energy for the derived value of $\delta = 0.7$ in the 
sub-TeV and TeV energy region. Such a rise is not observed so far in all the experiments and this controversy needs
an explanation. Perhaps, the rise of the anisotropy expected for $\delta = 0.7$ is related to the fixed mass of 
the primary cosmic rays: protons, helium etc. and the rise of anisotropy for protons is compensated by the rising 
fraction of heavier nuclei with their higher isotropy in the observed total primary flux or with the opposite 
phase (Erlykin and Wolfendale, 2015c).   

The interesting consequence of our scenario is
the large value of the escape pathlength ($\lambda_{esc}$) derived from the B/C ratio at low rigidities.
 It indicates that the observed excess of secondaries is produced in the region of high gas density. The 
consequence of this feature is that our Single Source should be searched for among gamma-ray sources. Finally, we 
conclude that the new data on the B/C ratio and the Lithium rigidity spectrum give support to our Single Source 
Model.          

\vspace{2mm}

{\bf Acknowledgements} \\

The authors are grateful to The Kohn Foundation for the financial support. We also thank Professor Sam Ting who 
has agreed to our use of the AMS-02 data referred to in the text.

\end{document}